\documentclass[aps,prb,twocolumn,superscriptaddress]{revtex4}

\usepackage{amsmath}
\usepackage{amssymb}
\usepackage{wasysym}
\usepackage{psfrag}
\usepackage{tabularx}
\usepackage[dvips]{graphicx}
\usepackage{enumitem}
\usepackage{color}
\usepackage{comment}
\usepackage[update,prepend]{epstopdf}
\usepackage{xcolor}
\def\beq{\begin{equation}}
\def\eeq{\end{equation}}
\def\bea{\begin{eqnarray}}
\def\eea{\end{eqnarray}}

\def\trigup{\bigtriangleup}

\newcommand{\Sp}{\mathbf{S}}

\def\rv {{\bf r}}
\def\kv {{\bf k}}

\begin{document}

\title{Chiral phase transition and thermal Hall effect in  an anisotropic spin model on the kagome lattice}

\author{F.A. G\'omez Albarrac\'in}
\email[corresponding author: ]{albarrac@fisica.unlp.edu.ar}
\affiliation{Instituto de F\'isica de L\'iquidos y Sistemas Biol\'ogicos (IFLYSIB), UNLP-CONICET, La Plata, Argentina and Departamento de F\'isica, Facultad de Ciencias Exactas,
Universidad Nacional de La Plata, c.c. 16, suc. 4, 1900 La Plata, Argentina.}
\affiliation{Departamento de Ciencas B\'asicas, Facultad de Ingenier\'ia, UNLP, La Plata, Argentina}

\author{H. D. Rosales}
\email[ ]{rosales@fisica.unlp.edu.ar}
\affiliation{Instituto de F\'isica de L\'iquidos y Sistemas Biol\'ogicos (IFLYSIB), UNLP-CONICET, La Plata, Argentina and Departamento de F\'isica, Facultad de Ciencias Exactas,
Universidad Nacional de La Plata, c.c. 16, suc. 4, 1900 La Plata, Argentina.}
\affiliation{Departamento de Ciencas B\'asicas, Facultad de Ingenier\'ia, UNLP, La Plata, Argentina}
\author{P. Pujol}
\email[]{pierre.pujol@irsamc.ups-tlse.fr}
\affiliation{Laboratoire de Physique Theorique-IRSAMC, CNRS and Universit\'e de Toulouse, UPS, Toulouse, F-31062, France.}

\begin{abstract}
We present a study of the thermal Hall effect in the extended Heisenberg model with $XXZ$ anisotropy in the kagome lattice. This model has the particularity that, in the classical case, and for a broad region in parameter space, an external magnetic field induces a chiral symmetry breaking: the ground state is a doubly degenerate $q=0$ order with either positive or negative net chirality. 
Here, we focus on the effect of this chiral phase transition in the thermal Hall conductivity using  Linear-Spin-Waves theory. We explore the topology and calculate the Chern numbers   of the magnonic bands, obtaining a variety of topological phase transitions.  We also compute  the magnonic effect  to the  critical temperature associated with the chiral phase transition ($T_c^{SW}$).  
Our main result is that, the thermal Hall conductivity, which is null for $T>T_c^{SW}$, becomes non-zero as a consequence of the spontaneous chiral symmetry breaking at low temperatures. Therefore, we present a simple model where it is possible to ``switch'' on/off the thermal transport properties introducing a magnetic field and heating or cooling the system.
\end{abstract}

\maketitle

\section{Introduction}

 One of the most significant current discussions in condensed matter physics concerns the connection between non-trivial topological properties and transport phenomena in insulating magnets\cite{Tokura2019}. It has been at the heart of numerous experimental and theoretical studies, mainly because these types of materials are candidates for carriers of the spin information without dissipation from Joule heating but with good transport coherence. Recently, particular attention has been brought upon the magnon thermal Hall effect (THE) \cite{THEOnose2010,THENagaosa,Matsumoto2011,Matsumoto2011b,Murakami2017}, where the transverse heat current induced by introducing a longitudinal thermal gradient is carried by magnonic excitations. 

The magnon THE was predicted theoretically  and observed experimentally in materials such as the insulating  ferromagnet Lu$_2$V$_2$O$_7$ \cite{THEOnose2010}, which has a pyrochlore lattice and antisymmetric Dzyaloshinskii-Moriya (DM) interactions perpendicular to the vanadium bonds. Other ferromagnetic pyrochlore insulators  include Ho$_2$V$_2$O$_7$, and In$_2$Mn$_2$O$_7$ \cite{THEPRB2012}. It has also been measured in perovskites  La$_2$NiMnO$_6$ and YTiO \cite{THEPRB2012} and kagome magnets  Cu(1-3, bdc) \cite{THEkagomePRL}, $[$CaCu$_3$(OH)$_6$Cl$_2\cdot$0.6H2O$]$ \cite{THEkagomePRL2}.
Magnon transport has also been theoretically studied  in different topological structures and models  \cite{LaurelFiettePRL,Mook1,Mook2,Owerre,WangPRB2017,THENandini,THEsquare}, which include both chiral and coplanar \cite{LaurelFiettePRB,Coplanar} systems. Furthermore, this transport phenomena has even led to the proposition of devices to manipulate the spin wave current in what is called ``topological magnonics'' \cite{TopMag}.

In a previous work (Ref.~[\onlinecite{HallPierre}]), we presented an extended $XXZ$ antiferromagnetic model in the kagome lattice  with an emergent  ``spontaneous'' Chern insulator, where the net chirality can be controlled by an external magnetic field. There is  a hidden phase transition in terms of the scalar chirality that separates the high-temperature  phase from the chiral low-temperature phase holding two $q=0$ ground states with opposite net scalar chirality. In this paper, we explore the consequences of this chiral phase transition in the thermal Hall conductivity. Using the Linear-Spin-Waves (LSW) theory approach, we first calculate the Chern numbers of the magnonic bands, and show that there are several topological phase transitions induced by the microscopic parameters. Then, we calculate the scalar chirality obtained from LSW and compute the  the magnonic effect to the classical critical temperature. Finally, we present the effects of the chiral phase transition and the associated symmetry breaking in the thermal conductivity: a nul contribution for $T>T_c^{SW}$. We close with discussion and conclusions.\\

\section{Model and Non-interacting Magnons}

\begin{figure}[htb]
\includegraphics[origin=c,width=0.7\columnwidth]{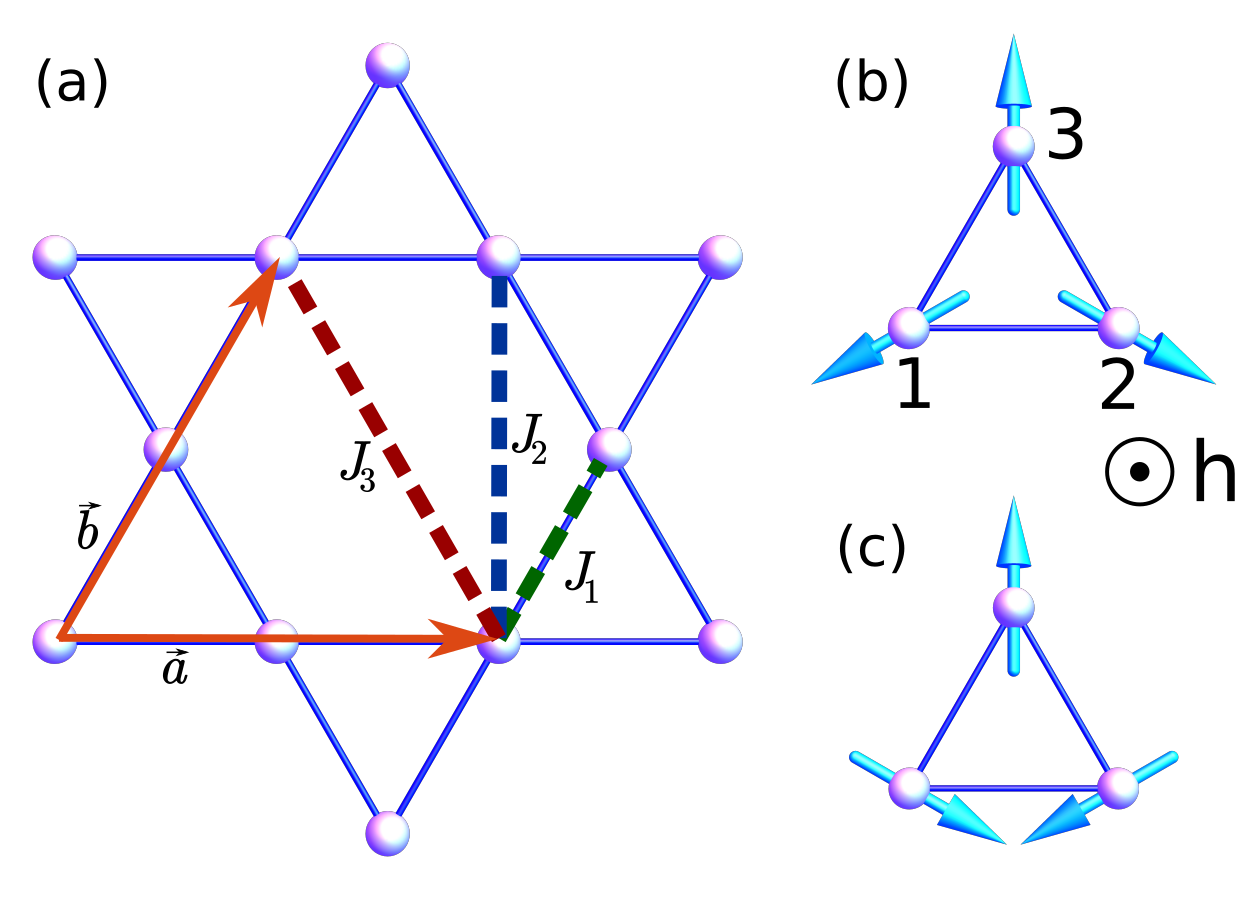}
\caption{\label{fig:latt}(a) The kagom\'e lattice with Bravais vectors $\vec{a}=(1,0)$ and $\vec{b}=(\frac{1}{2},\frac{\sqrt{3}}{2})$. 
First, second and third nearest neighbors exchange couplings, $J_1$, $J_2$ and  $J_3$ respectively, are indicated. We only consider $J_3$ as indicated in the figure.
(b-c) Two possible plaquette arrangements of the canted 120$^{\circ}$ ground state.} 

\end{figure}

We consider the extended antiferromagnetic Heisenberg model in the kagome lattice up to third nearest-neighbor interactions, taking only third nearest neighbors interactions across the hexagons (see Fig. \ref{fig:latt}).
\beq
H=\sum_{n=1}^3\sum_{\langle i,j\rangle_n} J_n\left(S^x_iS^x_j + S^y_iS^y_j + \Delta S^z_iS^z_j\right) - h\sum_iS^z_i
\label{eq:H}
\eeq
where $n$ indicates the $n-th$ nearest neighbor, $\Delta<1$ is the $XXZ$ anisotropy parameter and $h$ is the external magnetic field along the $z$ direction. In the SO(3) invariant $\Delta=1$ case, the classical $T=0$ phase diagram of this model is well known: it presents the so called ``cuboc'' phases (with  spontaneous and alternate scalar chirality), and a $q=0$ phase \cite{Lhuillier,FlaviaPierre} for $J_3<J_2<J_1$. At the special line $J_2=J_3<J_1$, the ground state has a semi-extensive degeneracy \cite{FlaviaPierre} where lines of spins from the $q=0$ order can be ``swapped''. For practical reasons, we will take $J_1=1$  for the rest of the manuscript.

 The combination of the $XXZ$ anisotropy and an external magnetic field induces a $q=0$ ``umbrella'' order with spontaneous non-zero net chirality. The emergence of scalar chirality in this simple model is quite remarkable, with a rich potential for unconventional phenomena.
The $xy$ projections of these two possible plaquette orders with opposite scalar chirality are shown in Fig. \ref{fig:latt} (b) and (c), where the three spins have the same projection along the field.

In these two possible ground states, the classical order is a   canted 120$^{\circ}$  plaquette. The state shown in Fig. \ref{fig:latt} (b) can be described  (minus a global rotation around the $z$ axis) as: $\vec{S}_1=S\left( -\frac{\sqrt{3}}{2}\sin\theta,-\frac{1}{2}\sin\theta,\cos\theta \right)$, $\vec{S}_2=S\left(\frac{\sqrt{3}}{2}\sin\theta,-\frac{1}{2}\sin\theta,\cos\theta \right)$ and $\vec{S}_3=S\left(0,\sin\theta,\cos\theta \right)$ where $S$ is the spin length and $\theta$ is the angle measured from the $z$ axis. 
As it is well known, the scalar chirality in a plaquette is defined as the triple product of the three spins $\chi^{0}_{\bigtriangleup}=\vec{S}_1\cdot(\vec{S}_2\times\vec{S}_3)$ which is a measure of the solid angle formed by them ($\chi^{0}_{\bigtriangleup}=S^3\,3/2\sqrt{3}\cos\theta\sin^2\theta$ for this configuration). 

In a recent  work \cite{HallPierre}, we focused on the special case $J_1=1,J_2=1/2,J_3=0$, $\Delta=0.9$ and showed that at low temperature the system undergoes a phase transition where the  lattice-only reflection symmetry $(x,y) \rightarrow (x,-y)$  is spontaneously broken.  As discussed in Ref. \onlinecite{FlaviaPierre}, this symmetry transforms $\chi_{\bigtriangleup,\bigtriangledown} \rightarrow -\chi_{\bigtriangledown,\bigtriangleup}$. Therefore, the relevant order parameter is in fact the total scalar chirality (per plaquette) $\chi_{tot}=\frac{1}{N_\bigtriangleup}\sum_{\bigtriangleup}\chi^{0}_{\bigtriangleup}$ where the sum involves all the triangular plaquettes $N_\bigtriangleup$. This will allow us to study the effect of  magnons in the critical temperature, defining the chirality operator, as we will show later.

In order to introduce  quantum spin fluctuations  and characterize the transport properties of the magnon excitations of this model, we resort to a linear spin wave (LSW) analysis\cite{AuerbachBook}.  Following the standard approach , we employ a three sublattice Holstein-Primakoff (HP) mapping with the bosonic operators (see Appendix for details).
\begin{figure*}
\includegraphics[width=2.1\columnwidth]{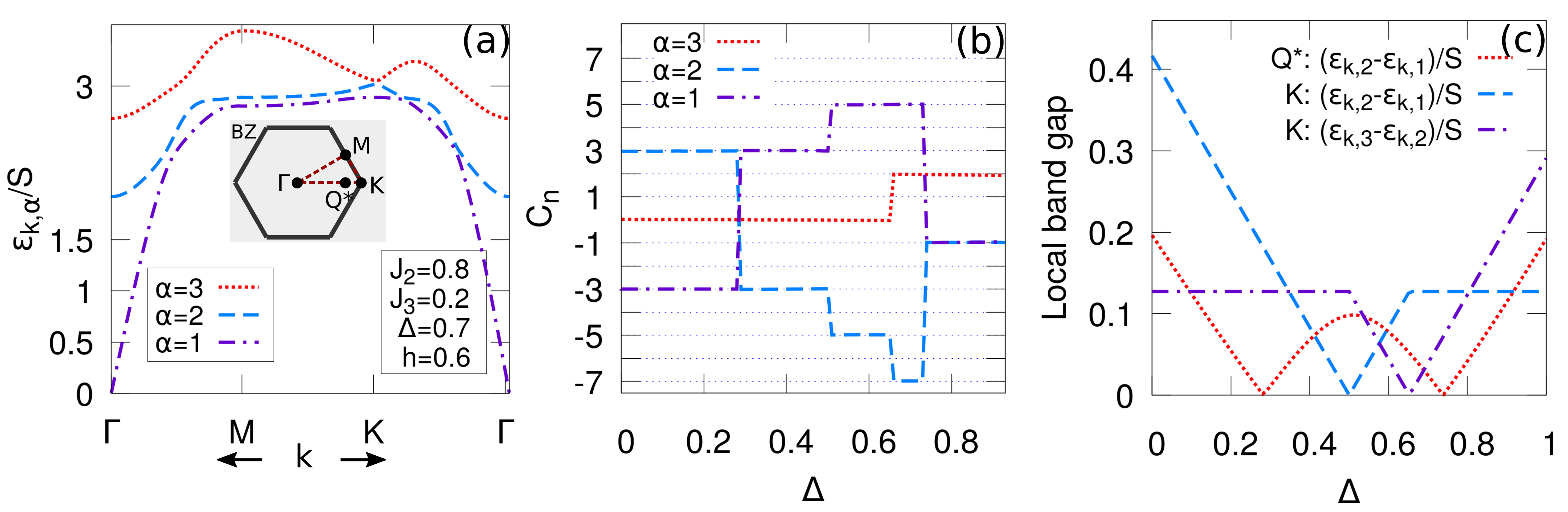}
\caption{\label{fig:ChernDelta} Fixing $J_2=0.8$, $J_3=0.2$ (a) Magnon spectrum for  $h/S=0.6$ , $\Delta=0.7$. The bands do not touch at any point in the BZ. (b) $C_n$ for each magnon band as a function of  $\Delta$, for $\theta=1.5$  (in radians)}. (c) Distance between magnon bands at different points in the BZ as a function of $\Delta$. When the distance is zero, we identify a closing of the local gap between bands.
\end{figure*}

Even though classically, there are only two types of solutions with opposite chirality,  there is a rich phenomenology in the magnon spectrum. To characterize the bands of the spectrum, we calculate the Chern number $C_n$ for each $n$-th band, defined as $ C_n=\frac{1}{2\pi}\int_{BZ}\Omega_n^z(\kv)dk^2$, where $\Omega_n^z(\kv)$ is the Berry curvature, $\Omega_n^z(\kv)=i\left\langle \frac{\partial u_n}{\partial\kv} \left|\times\right| \frac{\partial u_n}{\partial\kv}\right\rangle$, and $|u_n(\kv)\rangle$ are the Bloch waves in the $n$-th band. To calculate this quantity numerically, we resorted to the  efficient method detailed in Ref.[\onlinecite{ChernCalc}], taking up to $10000\times 10000$ points in the discretized Brillouin Zone (BZ). As we describe in the next subsection, depending on the microscopic parameters, even for small modifications of the classical solution, the associated magnonic bands present different $C_n$, which leads to several interesting phenomena.\\

\subsection{Topological magnonic bands and Topological Phase Transitions}

In order to discuss the magnon band structure, we choose as the classical groundstate one of the two $q=0$ states, shown in Fig. \ref{fig:latt}(b).  The magnon spectrum obtained from the LSW expansion is the same for both states, but the bands have opposite $C_n$. From a general analysis,  setting for example  $h/S\sim 0.6$ ($\theta\sim 1.5$), we find that there are regions in the parameter space ($J_2,J_3,\Delta$) with topologically non-trivial band structure  with different $C_n$. These regions are divided by topogical transitions, that occur when the magnon bands touch, and the ``local'' gap between them closes. For a representative case of this situation we fix $J_2=0.8$ and $J_3=0.2$ with $\Delta\in[0,0.9]$. The band structure for $\Delta =0.7$ is depicted in Fig. \ref{fig:ChernDelta}(a), where there are local gaps between all the bands;  $\epsilon_{\kv,\alpha}$ indicates the energy of the $\alpha=1,2,3$ band.

In  Fig. \ref{fig:ChernDelta}(b) we plot the Chern number of each band as a function of $\Delta$.   Since in the $\Delta \rightarrow 1$ limit the classical ground state is degenerate, we plot up to $\Delta=0.9$, to ensure an optimal numerical calculation of $C_n$.  The $C_n$ appear as steps in the constructed curve, and there are at least four topological transitions at different values of the anisotropy parameter. The sets of $C_n$ (from the lowest ($\epsilon_{\kv,1}$) to the top ($\epsilon_{\kv,3}$) band) go as $(-3,3,0)\rightarrow(3,-3,0)\rightarrow (5,-5,0)\rightarrow (5,-7,2) \rightarrow (-1,-1,2)$. As $\Delta$ increases, the topology of the bands change. For stronger $XXZ$ anisotropy, the lower bands have opposite $C_n$ and the top band has a trivial topology ($C_n=0$). For higher $\Delta \gtrsim 0.65$ the top band gets a non-trival $C_n$.In  Fig.  \ref{fig:ChernDelta}(c) we show the gap between successive bands
as a function of $\Delta$ for specific  points in the BZ.  The $Q^{*}$ point of the BZ mentioned in Fig. \ref{fig:ChernDelta}(c) is an incommensurate point between the $K$ and $\Gamma$ point, illustrated qualitatevely in the inset of Fig. \ref{fig:ChernDelta}(a). This value depends on the parameters, such as the external field and the anisotropy parameter. As expected, there is a perfect correspondence between the values of $\Delta$ where there is a topological transition  and the values of the anisotropy paramenter where two of the bands touch. There is also a correspondance between the magnitude of the jumps and the number of points in the BZ where the gap closes and reopens: in this particular example, there is a $\pm 6$ jump in the $C_n$ when the bands touch at the six $Q^{*}$, and a $\pm 2$ jump when they do at the $K(K')$ points. 

\begin{figure}
\includegraphics[width=0.7\columnwidth]{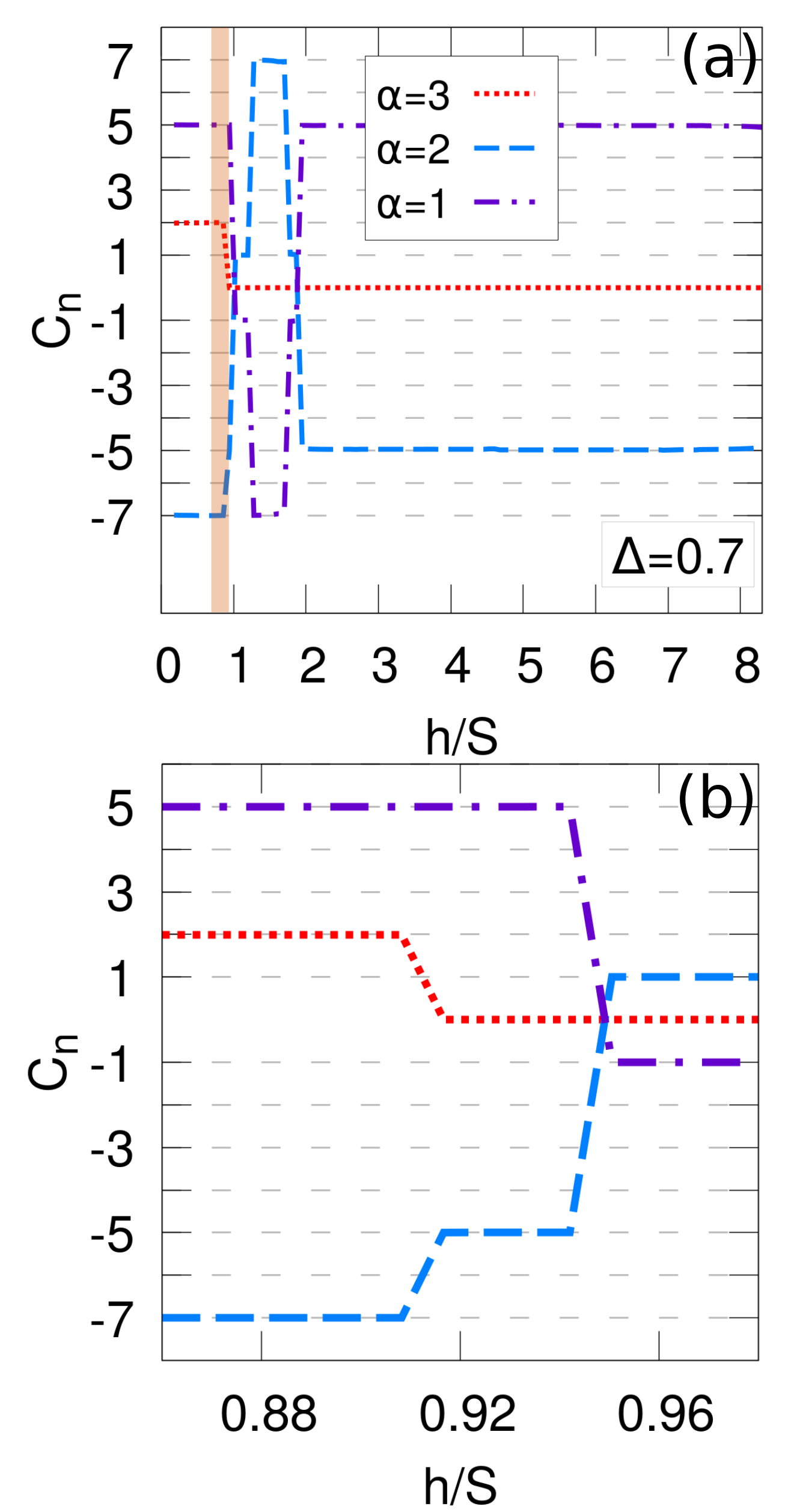}
\caption{\label{fig:ChernDeltaB} Fixing $J_2=0.8$, $J_3=0.2$, $\Delta=0.7$  (a) $C_n$ as a function of the external magnetic field $h/S$.  (b) Detail of the highlighted area from panel (a)}
\end{figure}

Another interesting issue is the role that the magnetic field plays on the topological transitions, even though classically the magnetic field just changes the canting angle of the spins. Varying the field triggers a series of topological transitions, that are reflected in the change of the $C_n$. We show this for $\Delta=0.7$ in Fig. \ref{fig:ChernDeltaB}(a), where the $C_n$ go  $(5,-7,2)\rightarrow (-5,5,0) \rightarrow (-1,1,0) \rightarrow (-7,7,0) \rightarrow (-1,1,0) \rightarrow (5,-5,0)$.  Since the classical saturation value is $h/S=8.52$ (as calculated according to Eq.(\ref{eq:hfield}) from the Appendix), we here plot up to $h/S=8.5$. The $(-5,5,0)$ intermediate region in the  $(5,-7,2)\rightarrow (-5,5,0) \rightarrow (-1,1,0)$ transition at low fields, highlighted in Fig. \ref{fig:ChernDeltaB}(a), is particularly narrow, and we zoom in this area in Fig. \ref{fig:ChernDeltaB}(b). As above, the surprisingly large values of the obtained $C_n$ are also remarkable. A similar feature was discussed in Ref.[\onlinecite{LaurelFiettePRB}], where this was attributed to an in-plane DM interaction. In our work, there are no antisymmetric interactions; the distinctive feature is the $XXZ$ anisotropy and the antiferromagnetic nature of the couplings. \\

\subsection{Chirality and Phase Transition} 

A key question in this work is the effect of  magnons in the classical phase transition and the consequences on the thermal transport properties. The relevant order parameter in this case,  since both ground states have the same canting angle, is not the magnetization but the scalar chirality,  which allows the detection of the spontaneous symmetry breaking at low temperature. To this end, we compute the quantum version of the scalar chirality $\chi_{tot}$ using HP tranformation and retaining terms up to quadratic order obtaining, 
\begin{eqnarray}
\langle\chi_{tot}\rangle&=&\left(1+\frac{3}{2\,S}\right)\chi^{0}_{\bigtriangleup}+\frac{S^2}{N_{\kv}}\sum^{3}_{\kv,\alpha=1}\left[\tilde{Q}^{\alpha\alpha}_{\kv}g(\epsilon_{\kv,\alpha})\right.\nonumber\\
&&\left.+\tilde{Q}^{\alpha+3,\alpha+3}_{\kv}(1+g(\epsilon_{\kv,\alpha}))\right]
\label{eq:chiQM}
\end{eqnarray}

\noindent where $N_{\kv}$ is the number of points in the Brillouin zone, $\chi^{0}_{\bigtriangleup}=S^3\frac{3\sqrt{3}}{2}\cos\theta\sin^2\theta$ is the classical scalar chirality for one triangular plaquette,  $g(\epsilon_{\kv,\alpha})$ is the Bose-Einstein distribution and $\tilde{Q}_{\kv}$ is the chirality operator matrix in the diagonal basis (explicit expressions in   the Appendix).

\begin{figure}[bh]
\includegraphics[width=0.99\columnwidth]{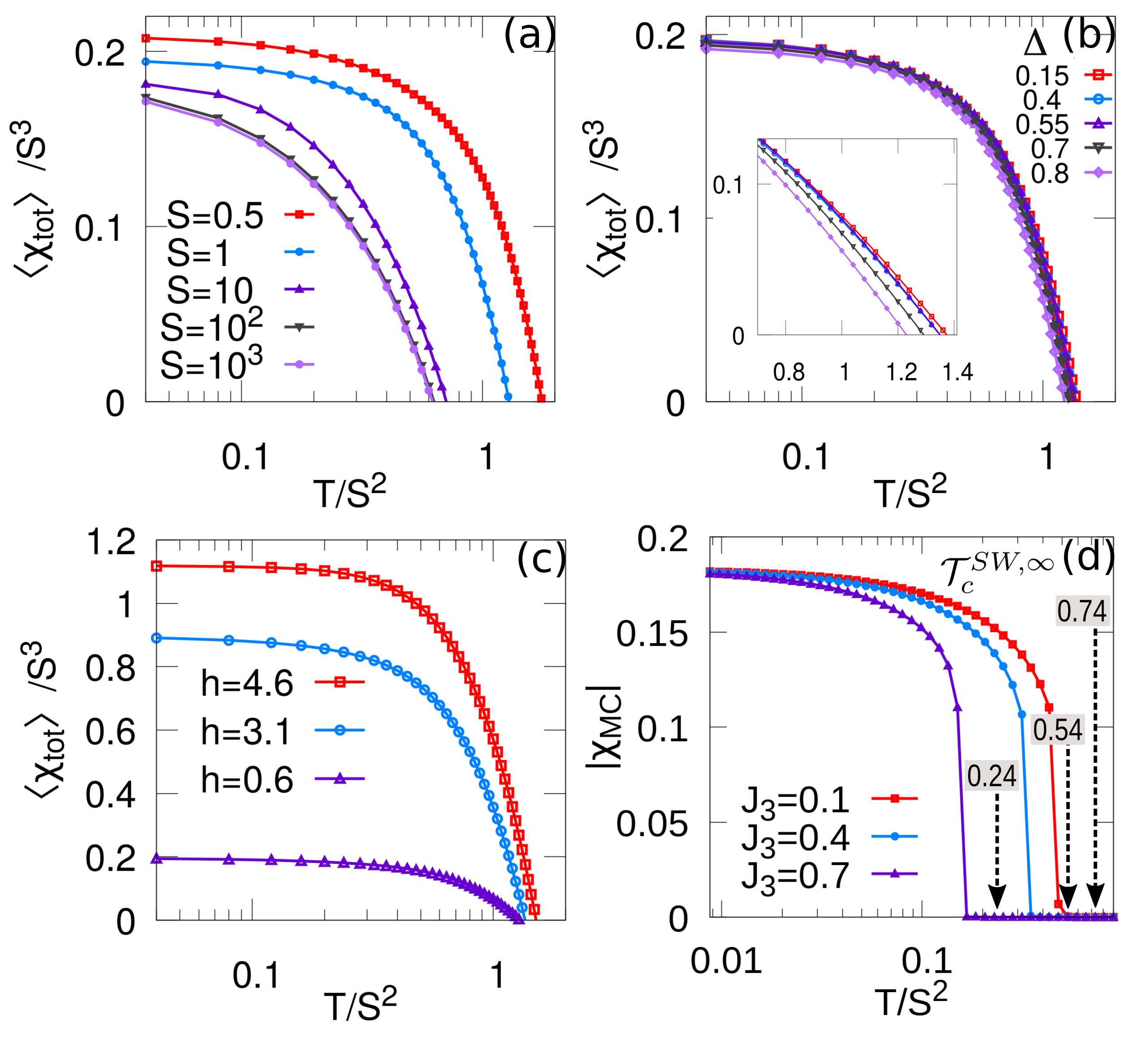}
\caption{\label{fig:ChiSW} Absolute value of the chirality parameter as a function of temperature  from LSW for $J_2=0.8$, $J_3=0.2$ and: (a) $\Delta=0.7$, $h/S=0.6$ and different values of  $S$; (b) $S=1$, $\theta=1.5$  and five values of $\Delta$, the inset zooms in to show that for higher $\Delta$, $T_c^{SW}$ is lower ; (c)  $S=1$, $\Delta=0.7$ and three values of $h$. (d) $|\chi_{\trigup}|$ vs. temperature  from MC simulations, $J_2=0.8$, $\Delta=0.7$, $h/S=0.6$ and  $J_3=0.1,0.4,0.7$. Vertical black arrows indicate the critical temperature $\mathcal{T}_c^{SW,\infty}$ obtained in the classical limit $S\rightarrow \infty$. Calculations where done for a discretization of $10000\times10000$ points in the BZ.}
\end{figure}

Let us first explore the dependence of critical temperature $T_c^{SW}$ in terms of the spin length $S$  taking  $J_2=0.8$, $J_3=0.2$, $\Delta=0.7$  $h/J_1=0.6$. From Fig. \ref{fig:ChiSW}(a),where we plot $\langle\chi_{tot}\rangle/S^3$ as a function of $T/S^2$ for different values of spin $S$,  we observe that  the chiral phase is stable up to the point $\langle\chi_{tot}\rangle/S^3=0$  which defines the critical temperature $T_c^{SW}$ for each $S$.    Looking at the $\langle\chi_{tot}\rangle/S^3$ vs $T/S^2$ curves for different values of $S$, it is clear that for a larger $S$ there is a smaller  $T_c^{SW}/S^2$. Moreover,  for the largest values of $S$, the curves tend to collapse around $T_c^{SW}/S^2\sim 0.34$. We show $\langle\chi_{tot}\rangle/S^3$  for $S=1$ as a function of temperature for different values of $\Delta$ (fixing $\theta=1.5$, Fig. \ref{fig:ChiSW}(b)) and external magnetic field (fixing $\Delta=0.7$, Fig. \ref{fig:ChiSW}(c)), where we can see that the behavior is robust. The inset in Fig \ref{fig:ChiSW}(b) shows that the critical temperature is lowered as the anisotropy parameter increases.

By Eq. (\ref{eq:chiQM}), we can compute in the classical limit $\mathcal{T}_c^{SW,\infty}=\lim_{S\rightarrow\infty} T_c^{SW}/S^2$,   defined by the condition $\langle\chi_{tot}\rangle=0$. After a simple analysis  (see Appendix for details) we obtain 

\begin{equation} 
\mathcal{T}_c^{SW,\infty}=-\frac{2\chi^{0}_{\Delta}}{S^3} \left[\frac{1}{N_{\kv}}\sum_\mathbf{k}\sum_{\alpha=1,2,3}\frac{\tilde{Q}^{\alpha\alpha}_{\kv}+\tilde{Q}^{\alpha+3,\alpha+3}_{\kv}}{\omega_{\kv,\alpha}}\right]^{-1}
\label{eq:Tc}
\end{equation}

\noindent where $2S\omega_{\kv,\alpha}=\epsilon_{\kv,\alpha}$.  Eq.(\ref{eq:Tc}) allows us to compare $\mathcal{T}_c^{SW,\infty}$  with Monte-Carlo (MC) simulations in Fig. \ref{fig:ChiSW}(c), as the system approaches the $J_2=J_3$ line, where the classical model has a semiextensive degeneracy. For the MC simulations, we resort to the Metropolis algorithm combined with overrelaxation (microcanonical) updates in system size of $3L^2$ sites ($L=30$). The estimated $\mathcal{T}_c^{SW,\infty}$, even within the LSW approximation, seems to be of the same order of magnitude as the one obtained form MC, a situation which contrasts with 3D systems with long range magnetic order, and for which in general LSW gives an huge overestimation of the critical temperature\cite{Li2018}. The reason for the good estimation of the critical temperature with LSW is likely to rely on the low value (compared to the microscopic parameters) of it, implying a low contribution of 
the terms proportional to $S^2$ in Eq.(\ref{eq:chiQM}),
The most important feature of our results is that lowering the temperature from the paramagnetic phase,  the quantum version of the model in Eq.(\ref{eq:H}) undergoes a phase transition in which the reflection symmetry is spontaneously broken. This has relevant consequences in the transport properties and the thermal Hall conductivity, which we will discuss below.\\

\subsection{Spontaneous Thermal Hall Conductivity}

The presence of a nontrivial Berry curvature in the magnon bands implies the existence of a thermal Hall signature provided that the Berry curvature is not odd in momentum. The thermal Hall conductivity $\kappa_{xy}$ may be calculated as \cite{THEPRB2014,Mook3}:

\begin{equation}
\kappa_{xy} = -\frac{k_B^2T}{(2\pi)^2\hbar}\sum_n\int_{BZ}\left[c_2[g(\epsilon_{\kv,\alpha})]-\frac{\pi^2}{3}\right]\Omega_n^z(\vec{k})d^2k
\end{equation}
where $k_B$ is the Boltzmann constant, $g(\epsilon_{\kv,\alpha})$ is the Bose-Einstein distribution  and $c_2$ is defined as:

\begin{equation}
 c_2(x)=(1+x)\left[\ln\left(\frac{1+x}{x}\right) \right]^2-(\ln x)^2-2\mathtt{Li}_2(-x)
\end{equation}
where $\mathtt{Li}_2(x)$ is the dilogarithm.

\begin{figure}[h!]
\includegraphics[width=0.7\columnwidth]{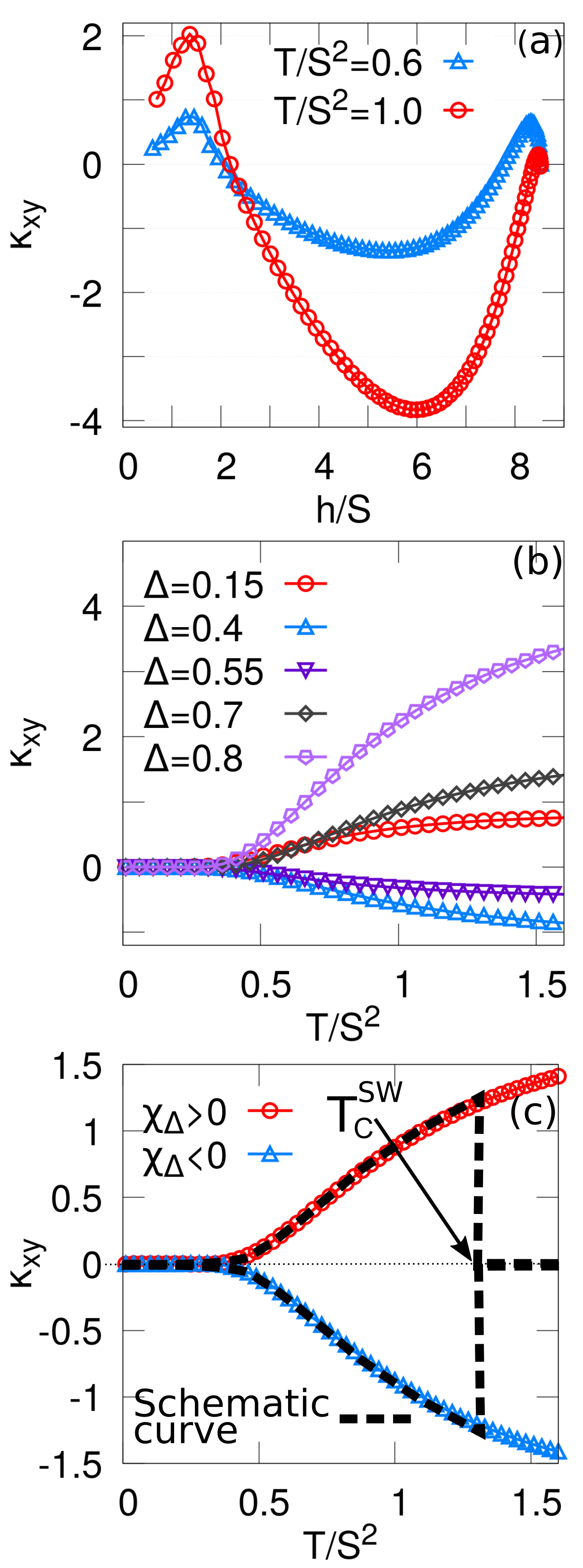}
\caption{\label{fig:Kappa2Chi} For $S=1$, $J_2=0.8$, $J_3=0.2$  (a) $\kappa_{xy}$ as a function of $h$ taking $\Delta=0.7$ for two different temperatures $T=0.6$ and $T=1$; (b) thermal conductivity $\kappa_{xy}$ as a function of temperature for different values of $\Delta$, for $\theta=1.5$; (c) $\kappa_{xy}$ as a function of temperature obtained from the magnon spectrum of the two possible classical ground states with opposite scalar chirality for $\Delta=0.7, h/S=0.6$.}
\end{figure}

In Fig.\ref{fig:Kappa2Chi}(a) we show $\kappa_{xy}$ as a function of the external field for  two temperatures $T<T_c^{SW}$.  In Fig. \ref{fig:Kappa2Chi}(b) we show $\kappa_{xy}$ (in units of $k_B^2/(2\pi\hbar)$) for $S=1$ as a function of the temperature for  different $\Delta$, corresponding to different  regions marked by the topological transitions in Fig. \ref{fig:ChernDelta}(b).

The most important feature of this study is shown in Fig. \ref{fig:Kappa2Chi}(c). This figure shows that the sign of $\kappa_{xy}$ depends on the sign of the scalar chirality from the classical ground state. This suggests that $\kappa_{xy}$ will have a spontaneos sign for $T<T_c^{SW}$. However, because $\langle\chi_{tot}\rangle=0$ for  $T>T_c^{SW}$,  $\kappa_{xy}$ must vanish for $T>T_c^{SW}$, implying a ``switchable'' THE. Unfortunately, this suppression of the  $\kappa_{xy}$ is not completely captured by LSW. We expect magnon interactions \cite{Li2018,Zhitomirsky}, not included at this stage, to establish the cancellation of $\kappa_{xy}$  for $T>T_c^{SW}$. This requires a higher order spin wave calculation (in powers of $1/S$) which is beyond the scope of this paper. Nonetheless, given the previous discussion, we propose a cutoff in  $\kappa_{xy}$ as  represented in Fig. \ref{fig:Kappa2Chi}(c) with a dashed black line. \\

\section{Discussion and Conclusions}

The purpose of the present work was to study a chiral phase transition in an anisotropic Heisenberg model in the kagome lattice, in which the reflection symmetry is spontaneously broken, and where the low temperature phase shows a thermal Hall effect.

To this end, using the Linear-Spin-Waves theory approach, we have studied numerically the topology of the magnonic bands and their  Chern numbers. We have shown that there are several topological transitions driven by the magnetic field and the microscopic parameters. We have also studied the effect of magnons  in the chiral phase transition in terms of these parameters. We have paid a particular attention to the dependence of the critical temperature with the magnitude of the spins and obtained an interesting result comparing the value obtained taking the  classical limit ($S\to\infty$) with Monte Carlo simulations. 

Finally, we calculate the thermal Hall conductivity as a function of temperature showing that effectively its sign is ``spontaneous''.  Therefore, we show with this simple model that it is possible to ``switch'' on  the thermal transport properties by manipulation of the external parameters.

A unique feature of this work resides in the chiral phase transition. Previous works have established that a non-zero scalar chirality in the ground state could lead to THE. Here, the chirality  serves as an order parameter associated with a spontaneous broken symmetry, and thus allows us to conjecture the behaviour of the thermal Hall conductivity with temperature and define a critical temperature above which the thermal Hall conductivity is suppressed. In addition, we have shown that this conductivity can be tuned by the external magnetic field and the $XXZ$ anisotropy. We would like to point out the fact that, although the structure of the kagome lattice allows for Dzyaloshinskii-Moriya interactions, there is still room for systems in which a spontaneously broken symmetry chiral phase as the one observed here is present (see Ref. [\onlinecite{FlaviaPierre}] for a more complicated example). In this sense the example studied here has to be though as the simplest of a family of systems presenting a low temperature phase with a spontaneous THE.

Following the previous discussion, it would be interesting to consider magnon interactions going beyond LSW theory including three and four bosonic terms to obtain a more accurate estimation of the conductivity near the chiral phase transition. We defer this for future investigations.

\section*{Acknowledgments}
We would like to thank the ``Laboratoire International Associ\'e''  LIA LICOQ for support and 
Mike Zhitomirsky for very fruitful discussions. 
H.D.R. and F.A.G.A. acknowledge the Laboratoire de Physique Th\'eorique
(LPT) in Toulouse for their hospitality during their 2019 visits.
H.D.R. and F.A.G.A. are partially supported by PIP 2015-0813 CONICET and SECyT-UNLP. 
H.D.R. acknowledges support from PICT 2016-4083 and F.A.G.A. from PICT 2018-02968


\section*{APPENDIX  - SPIN WAVE THEORY}

\subsection{Quadratic bosonic Hamiltonian and magnonic spectrum}
We consider a kagome-lattice antiferromagnet with anisotropic XXZ exchange interactions
up to third nearest-neighbors, taking only third nearest neighbors interactions across the hexagons. First we write the Hamiltonian as 

\begin{figure}[htb]
\includegraphics[origin=c,width=0.8\columnwidth]{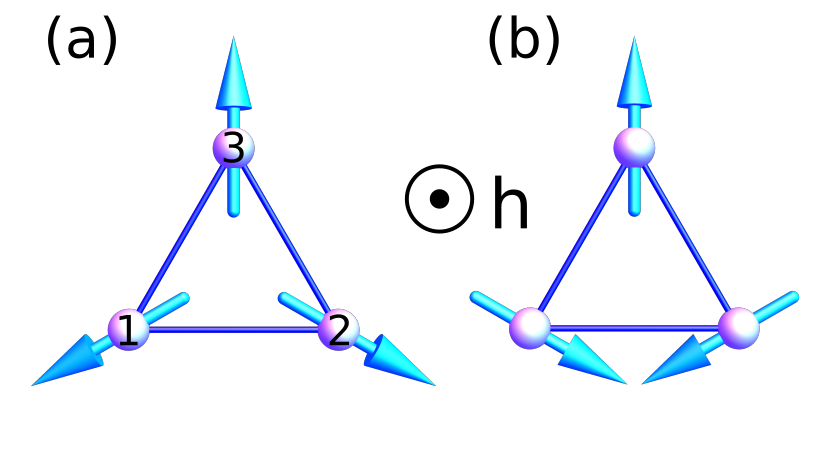}
\caption{\label{fig:suppFig1} Classical configuration for one plaquette: with  (a) positive and (b) negative scalar chirality.  The numbers 1,2 and 3 indicate spins from the three different triangular sublattices from the kagome lattice.} 

\end{figure}

\begin{widetext}
\bea
H=\sum_{n=1}^3\sum_{\langle i,\alpha;j,\beta\rangle_n} J_n\left(S^x_{i,\alpha}S^x_{j,\beta} + S^y_{i,\alpha}S^y_{i,\beta} + \Delta S^z_{i,\alpha}S^z_{j,\beta}\right) - h\sum_{i,\alpha}S^z_{i,\alpha}
\label{eq:Hboson}
\eea
\end{widetext}

\noindent where $i$ and $j$ are indices for the positions $\rv_i,\rv_j$ in the periodic Bravais lattice, $\alpha$ and $\beta$ are sublattice indices  ($1, 2$ ,or $3$ as indicated in Fig. \ref{fig:suppFig1}), $n$ indicates the $n$-th nearest neighbor, $\Delta < 1$ is the XXZ anisotropy parameter and $h$ is the external magnetic field along the $z$ direction. For $J_3 < J_2 < J_1$ and $h>0$, the classical ground state correspond to spins forming an ``umbrella''  $q = 0$  order with spontaneous non-zero net chirality and  two possible $xy$ projections, shown in Fig. \ref{fig:suppFig1}. The classical energy per plaquette is given by:
\begin{widetext}
\bea
\frac{\mathcal{E}_0}{N_{\bigtriangleup}}&=&\frac{S^2}{2}\left[6 (2 J_1 + 2 J_2 + J_3) \Delta \cos^2\theta -     6 (J_1 + J_2 - J_3) \sin^2\theta\right]-3\, h\,S \cos\theta,
\eea
\end{widetext}
where $N_{\bigtriangleup}$ is the total number of plaquettes, $\theta$ is the canting angle between the spins and the magnetic field. Minimization of the classical energy fixes  $\theta$  to
\bea \label{eq:hfield}
h&=&S\,\left[(2 + 4\,\Delta) (J_1 + J_2) +    2 J_3 (\Delta- 1)\right] \cos\theta
\eea

In order to include quantum spin fluctuations about the classical magnetic order (Fig. \ref{fig:suppFig1}(a) or (b)), we construct the Hamiltonian of free spin waves using the Holstein-Primakoff (HP)  transformation\cite{AuerbachBook}. We follow the conventional strategy of the large-$S$ approach, we consider three sublattices, indicated in Fig. \ref{fig:suppFig1} (a), and define a local coordinate system along the direction of the classical ground state as spin quantization axis $\{x'_n,y'_n,z'_n\}$. The HP transformation in the local frame reads

\bea \label{eq:HP}
\tilde{S}_{i,\alpha}^z&=&S-a_{i,\alpha}^{\dagger}a_{i,\alpha}\\ 
\tilde{S}_{i,\alpha}^+&=&\sqrt{2S-a_{i,\alpha}^{\dagger}a_{i,\alpha}}a_{i,\alpha}\approx a_{i,\alpha}\sqrt{2S}\\
\tilde{S}_{i,\alpha}^-&=&a_{i,\alpha}^{\dagger}\sqrt{2S-a_{i,\alpha}^{\dagger}a_{i,\alpha}}\approx a^{\dagger}_{i,\alpha}\sqrt{2S}
\eea

with $a_{i,\alpha}^{\dagger}$ ($a_{i,\alpha}$) being a bosonic creation (annihilation) operator of the sublattice $\alpha$ at the cell $i$. 
Using this mapping in the Hamiltonian given by Eq. (\ref{eq:Hboson}) and after a Fourier transformation $a_{i,\alpha}=(1/\sqrt{N})\sum_{\kv}e^{i\kv\cdot(\rv_i+\rv_\alpha)}a_{\kv,\alpha}$, where $N$ is the total number of unit cells, $\rv$ denotes the position of the unit cell and $\rv_\alpha$ are the internal  positions of the sublattices. 
The bilinear bosonic  Hamiltonian reads
\begin{eqnarray}
H_{2}&=&S\sum_{\kv}\Psi^{\dagger}_{\kv}\cdot M_{\kv}\cdot\Psi_{\kv}
\end{eqnarray}
\noindent where $\Psi^{\dagger}_{\kv}=\{a^{\dagger}_{\kv,1},a^{\dagger}_{\kv,2},a^{\dagger}_{\kv,3},a_{-\kv,1},a_{-\kv,2},a_{-\kv,3}\}$ and the matrix  $M_{\kv}$ is 

\bea
\text{M}_\kv=\left[ 
\begin{array}{cc}
D_{\kv} & C_{\kv}\\
C^{\dagger}_{\kv} & D^{T}_{-\kv} 
\end{array} \right]
\label{eq:matrixK}
\eea
\noindent where the sub matrix $D_{\kv}, C_{\kv}$ have elements 

\begin{widetext}
\bea
D_{\kv}^{11}&=&\frac{1}{2} (h/S\,c\theta-2\,c\theta^2 \Delta (2 J_1+2 J_2+J_3)+2 (J_1+J_2-J_3) s\theta^2+J_3 (1+c\theta^2+\Delta\,s\theta^2) \cos(2 (q_1-q_2)))\nonumber\\
D_{\kv}^{21}&=&-\frac{1}{8} (3+c2\theta+4 i \sqrt{3}\, c\theta-2 \Delta+2\, c2\theta\, \Delta) (J_1 \cos( q_1)+J_2 \cos(q_1-2 q_2))\nonumber\\
D_{\kv}^{31}&=&\frac{1}{4} (-1+2 i \sqrt{3}\, c\theta-c\theta^2+2\, \Delta\, s\theta^2) (J_2 \cos(2 q_1-q_2)+J_1 \cos(q_2))\nonumber\\
D_{\kv}^{12}&=&-\frac{1}{8} (3+c2\theta-4 i \sqrt{3} c\theta-2\,\Delta+2\,c2\theta\,\Delta) (J_1 \cos( q_1)+J_2 \cos(q_1-2 q_2))\nonumber\\
D_{\kv}^{22}&=&\frac{1}{2} (h/S\,c\theta-2 c\theta^2\,\Delta (2 J_1+2 J_2+J_3)+2 (J_1+J_2-J_3) s\theta^2+J_3 (1+c\theta^2+\Delta\,s\theta^2) \cos(2 q_2))\nonumber\\
D_{\kv}^{32}&=&\frac{1}{4} (-1-2 i \sqrt{3} c\theta-c\theta^2+2\,\Delta\,s\theta^2) (J_1 \cos(q_1-q_2)+J_2 \cos(q_1+q_2))\nonumber\\
D_{\kv}^{13}&=&\frac{1}{4} (-1-2 i \sqrt{3} c\theta-c\theta^2+2\,\Delta\,s\theta^2) (J_2 \cos(2 q_1-q_2)+J_1 \cos(q_2))\nonumber\\
D_{\kv}^{23}&=&\frac{1}{4} (-1+2 i \sqrt{3} c\theta-c\theta^2+2\,\Delta\,s\theta^2) (J_1 \cos(q_1-q_2)+J_2 \cos(q_1+q_2))\nonumber\\
D_{\kv}^{33}&=&\frac{1}{2} (h/S\,c\theta -2 c\theta^2\,\Delta (2 J_1+2 J_2+J_3)+2 (J_1+J_2-J_3) s\theta^2+J_3 (1+c\theta^2+\Delta s\theta^2) \cos(2 q_1))\nonumber
\label{eq:matrixD}
\eea
\end{widetext}
\bea
C_{\kv}^{11}&=&\frac{1}{2} (-1+\Delta) J_3 s\theta^2 \cos(2 (q_1-q_2))\nonumber\\
C_{\kv}^{21}&=&\frac{1}{4} (1+2 \Delta) s\theta^2 (J_1 \cos(q_1)+J_2 \cos(q_1-2 q_2))\nonumber\\
C_{\kv}^{31}&=&\frac{1}{4} (1+2 \Delta) s\theta^2 (J_2 \cos(2 q_1-q_2)+J_1 \cos(q_2))\nonumber\\
C_{\kv}^{12}&=&\frac{1}{4} (1+2 \Delta) s\theta^2 (J_1 \cos(q_1)+J_2 \cos(q_1-2 q_2))\nonumber\\
C_{\kv}^{22}&=&\frac{1}{2} (-1+\Delta) J_3 s\theta^2 \cos(2 q_2)\nonumber\\
C_{\kv}^{32}&=&\frac{1}{4} (1+2 \Delta) s\theta^2 (J_1 \cos(q_1-q_2)+J_2 \cos(q_1+q_2))\nonumber\\
C_{\kv}^{13}&=&\frac{1}{4} (1+2 \Delta) s\theta^2 (J_2 \cos(2 q_1-q_2)+J_1 \cos(q_2))\nonumber\\
C_{\kv}^{23}&=&\frac{1}{4} (1+2 \Delta) s\theta^2 (J_1 \cos(q_1-q_2)+J_2 \cos(q_1+q_2))\nonumber\\
C_{\kv}^{33}&=&\frac{1}{2} (-1+\Delta) J_3 s\theta^2 \cos(2 q_1)\nonumber
\label{eq:matrixC}
\eea
\noindent where have defined $q_1=k_x$, $q_2=k_x/2+\sqrt{3}/2k_y$, $s\theta=\sin\theta$, $c\theta=\cos\theta$, $s2\theta=\sin2\theta$, $c2\theta=\cos2\theta$.

To diagonalize the bosonic Hamiltonian we perform a paraunitary Bogoliubov transformation\cite{Colpa78}  $M_d = U^{\dagger}_{\kv}\cdot M_{\kv}\cdot U_{\kv}=\text{diag}\{w_{\kv,1},w_{\kv,2},w_{\kv,3},w_{-\kv,1},w_{-\kv,2},w_{-\kv,3}\}$ with $U^{\dagger}_{\kv}\cdot\sigma_3\cdot U_{\kv}=\sigma_3$, $\sigma_3=\text{diag}\{1,1,1,-1,-1,-1\}$. The bilinear Hamiltonian becomes

\begin{eqnarray}
H_{eff}&=&\mathcal{E}_0+\sum_{\kv,\alpha}\epsilon_{\kv,\alpha}\,\left(b^{\dagger}_{\kv,\alpha}\,b_{\kv,\alpha}+\frac{1}{2}\right)
\end{eqnarray}

\noindent where the energy of the magnon bands is $\epsilon_{\kv,\alpha}=2S\omega_{\kv,\alpha}$ \cite{Zhitomirsky}
%

%
\subsection{Chirality order parameter}

Following the same strategy, we perform the HP transformation (Eq.(\ref{eq:HP})) in the chirality operator $\chi_{tot}=\frac{1}{N_\bigtriangleup}\sum_{\bigtriangleup}\chi_{\bigtriangleup}$, with $\chi_{\bigtriangleup}=\Sp_i\cdot(\Sp_j\times\Sp_k)$. Retaining up to quadratic bosonic terms, we obtain
\begin{widetext}
\bea
\langle\chi_{tot}\rangle&=&\left(S^3+\frac{3}{2}S^2\right)\frac{3\sqrt{3}}{2}\cos\theta\sin^2\theta+\frac{S^2}{N_{\kv}}\sum^{3}_{\kv\alpha=1}\left[\tilde{Q}^{\alpha\alpha}_{\kv}g(\epsilon_{\kv,\alpha})+\tilde{Q}^{\alpha+3,\alpha+3}_{\kv}(1+g(\epsilon_{\kv,\alpha}))\right]
\label{eq:chiQMapp}
\eea
\end{widetext}

where $N_{\kv}$ is the number of points in the Brillouin zone, $g(\epsilon_{\kv,\alpha})$ is the Bose-Einstein distribution and $\tilde{Q}_{\kv}=U^{\dagger}_{\kv}\cdot Q\cdot U_{\kv}$ with

\begin{widetext}
\bea
Q=\left[ 
\begin{array}{cccccc}
-B & A& A^{*}& 0& -\frac{B}{2}& -\frac{B}{2}\\
A^{*}& -B& A& -\frac{B}{2}& 0& -\frac{B}{2}\\
A& A^{*}& -B& -\frac{B}{2}& -\frac{B}{2}& 0\\
0& -\frac{B}{2}& -\frac{B}{2}& -B& A^{*}& A\\
-\frac{B}{2}& 0& -\frac{B}{2}& A& -B& A^{*}\\
-\frac{B}{2}& -\frac{B}{2}& 0& A^{*}& A& -B
\end{array} \right]
\label{eq:matrixQ}
\eea
\end{widetext}
with $A=\frac{1}{32}(8\,i+5\sqrt{3}\cos\theta+3\sqrt{3}\cos3\theta$ and $B=\frac{3\sqrt{3}}{4}\cos\theta\sin^2\theta$.

 From Eq.(\ref{eq:chiQMapp})  we can obtain the value of the critical temperature in the classical limit, $\mathcal{T}_c^{SW,\infty}=\lim_{S\rightarrow\infty} T_c^{SW}/S^2$. In order to do this, we  take the $S\rightarrow\infty$ limit in the Bose distribution function (setting the Boltzmann constant $k_B=1$):

\begin{eqnarray}
g(\epsilon_{\kv,\alpha})&=&g(2S\omega_{\kv,\alpha})=\frac{1}{\exp^{\frac{2S\omega_{\kv,\alpha}}{T}}-1} = \frac{1}{\exp^{\frac{2\omega_{\kv,\alpha}}{S}\frac{1}{T/S^2}}-1}  \nonumber \\
\Longrightarrow \lim_{S\to \infty} g(2S\omega_{\kv,\alpha}) &\sim& \frac{1}{1 + \frac{2\omega_{\kv,\alpha}}{S}\frac{1}{T/S^2} + ... -1}    = \frac{S\mathcal{T}_c^{SW,\infty}}{2\omega_{\kv,\alpha}}
\end{eqnarray}

Inserting the above expression for $g(\epsilon_{\kv,\alpha})$ in Eq.(\ref{eq:chiQMapp}) and taking $\langle \chi_{tot} \rangle = 0$ we obtain the expression for $\mathcal{T}_c^{SW,\infty}$ shown in  Eq.(\ref{eq:Tc}) in the main text.

\end{document}